\documentclass[12pt]{article}
\usepackage{amsmath,amssymb,epsfig}
\makeatletter \@addtoreset{equation}{section} \makeatother
\renewcommand{\theequation}{\thesection.\arabic{equation}}
\newcommand{\ba}{\begin{array}}
\newcommand{\ea}{\end{array}}
\newcommand{\beq}{\begin{equation}}
\newcommand{\eeq}{\end{equation}}
\newcommand{\bea}{\begin{eqnarray}}
\newcommand{\eea}{\end{eqnarray}}

\def\bce{\begin{center}}
\def\ece{\end{center}}

\def\nonu{\nonumber}

\def\pa{\partial}

\def\be{\beta}

\newcommand{\tr}{\mbox{Tr}}

\def\eps6{{\displaystyle \mathop{\epsilon}^{6}}{}}

\def\nab6{{\displaystyle \mathop{\nabla}^{6}}{}}

\def\0{{\sst{(0)}}}
\def\1{{\sst{(1)}}}
\def\2{{\sst{(2)}}}
\def\3{{\sst{(3)}}}
\def\4{{\sst{(4)}}}
\def\5{{\sst{(5)}}}
\def\6{{\sst{(6)}}}
\def\7{{\sst{(7)}}}
\def\8{{\sst{(8)}}}

\def\ba{\begin{array}}
\def\ea{\end{array}}
\def\beq{\begin{equation}}
\def\eeq{\end{equation}}
\def\be{\begin{equation}}
\def\ee{\end{equation}}

\def\tr{\mathop{\rm tr}}

\def\eps{\epsilon}

\def\ba{\begin{array}}
\def\ea{\end{array}}
\def\beq{\begin{equation}}
\def\eeq{\end{equation}}
\def\be{\begin{equation}}
\def\ee{\end{equation}}

\def\tr{\mathop{\rm tr}}

\def\eps{\epsilon}

\newcommand{\bean}{\begin{eqnarray*}}
\newcommand{\eean}{\end{eqnarray*}}

\begin{document}
\thispagestyle{empty} \addtocounter{page}{-1}
\begin{flushright}
{\tt hep-th/0610025}\\
\end{flushright}

\vspace*{1.3cm}

\centerline{ \Large \bf M-theory Lift of Meta-Stable Brane Configuration}
\vspace{.3cm} 
\centerline{ \Large \bf  in Symplectic and Orthogonal Gauge Groups} 
\vspace*{1.5cm}
\centerline{{\bf Changhyun Ahn} 
} 
\vspace*{1.0cm} 
\centerline{\it 
Department of Physics, Kyungpook National University, Taegu
702-701, Korea} 
\vspace*{0.8cm} 
\centerline{\tt ahn@knu.ac.kr} 
\vskip2cm

\centerline{\bf Abstract}
\vspace*{0.5cm}

The M-theory lift for the supersymmetry breaking IIA brane configuration 
corresponding to the meta-stable state of ${\cal N}=1$ unitary
supersymmetric Yang-Mills theory with massive flavors 
was found by Bena et al(hep-th/0608157) recently.
We extend this to 
symplectic and orthogonal gauge groups by analyzing the previously 
known results
on M-theory lifts of supersymmetric IIA brane configurations.

\baselineskip=18pt
\newpage
\renewcommand{\theequation}
{\arabic{section}\mbox{.}\arabic{equation}}

\section{Introduction}

In the ``electric theory'', the standard brane 
configuration \cite{EGK,GK} consists of $N_c$ D4-branes stretched 
between NS5-brane and NS5'-brane along the $x^6$ direction,
with $N_f$ D6-branes to the left of the NS5-brane, each of which is 
connected to the NS5-brane by a D4-brane. From this, 
one obtains ${\cal N}=1$ $SU(N_c)$ supersymmetric Yang-Mills theory
with massless $N_f$ flavors by taking a string coupling,
a string scale and a distance between NS5-brane and NS5'-brane to be 
zero
$g_s, \ell_s, \Delta L \rightarrow 0$ with fixed gauge coupling where
$\frac{1}{g_{elec}^2}=\frac{|\Delta L|}{g_s \ell_s}$(or equivalently
by removing all the 
massive states).
 
In the ``magnetic theory'', 
the standard brane 
configuration \cite{EGK,GK} consists of $(N_f-N_c)$ D4-branes stretched 
between NS5'-brane and NS5-brane along the $x^6$ direction,
with $N_f$ D6-branes to the left of the NS5'-brane, each of which is 
connected to the NS5'-brane by a D4-brane.
From this,
one obtains ${\cal N}=1$ $SU(N_f-N_c)$ supersymmetric Yang-Mills theory
with massless $N_f$ flavors by taking 
$g_s, \ell_s, \Delta L, L_0 \rightarrow 0$ where $L_0$ is a distance 
between D6-branes and NS5'-brane
with fixed gauge coupling where
$\frac{1}{g_{mag}^2}=\frac{\Delta L}{g_s \ell_s}$ and Yukawa coupling
$h=\sqrt{\frac{g_s \ell_s}{L_0}}$ in the notation of \cite{BGHSS}.

We can turn on a mass matrix for the quarks in the ``electric theory''
by adding 
a superpotential $\tr m Q \widetilde{Q}$. In the brane configuration,
masses correspond to relative displacement of the D6- and D4-branes in the 
45 directions and  
the relation between mass $m$ and this relative displacement $\Delta x$
is given by $m =\frac{\Delta x}{\ell_s^2}$. The corresponding 
field theory ``decoupling limit'' or ``SQCD limit'' defined in \cite{BGHSS} 
can be done in this case.

Similarly, in the ``magnetic theory'', one can deform the magnetic 
superpotential $M q \widetilde{q}$ 
by a linear term in $M$ where $M$ is magnetic meson fields.
This linear term in the deformed superpotential breaks supersymmetry
at tree level(and to all orders in perturbation theory) by the rank 
condition \cite{ISS}.  In this magnetic phase, the minimal-energy 
supersymmetry breaking 
brane configuration \cite{OO1,FGU,BGHSS} 
is the analogue of the supersymmetry breaking 
vacuum found in the magnetic dual to SQCD at tree level, although 
this brane configuration 
is not related to the meta-stable state of SQCD \cite{BGHSS}. 

M-theory lifts of the supersymmetric brane configurations 
can be described by M5-branes wrapping holomorphic curves
in Taub-NUT multiplied by flat two dimensions. Both the electric 
and magnetic brane configurations at zero mass 
lift the same holomorphic curves
in M-theory.
That is, two different brane setups are described by the same M-theory 
configuration \cite{SS,Hori} and are connected to each other 
through electric-magnetic duality.
However, for nonzero mass, the only one component M-theory curve 
reduces to the 
supersymmetric electric brane configuration 
as $R \rightarrow 0$ where $R$ is a radius of eleventh direction, 
as observed in \cite{BGHSS}.
There is no smooth limit which corresponds to a magnetic brane 
configuration.

Recently, the lifting of the tree-level supersymmetry breaking 
brane configuration was studied in \cite{BGHSS} 
by computing the equations of motion 
for the minimal area nonholomorphic 
curves in  Taub-NUT multiplied by flat two dimensions.
It turns out that there is no meta-stable brane configuration in the 
``D-brane limit'' defined in \cite{BGHSS} 
of MQCD when we consider string interactions. 
The behavior at infinity of this nonsupersymmetric brane configuration 
is different from that of the standard 
supersymmetric ground state of MQCD. 

In this paper, we generalize the work of \cite{BGHSS} to the
meta-stable brane configurations in 
symplectic
and orthogonal gauge groups by recalling the known results 
\cite{AOT,AOT1,CS} 
on M-theory lifts of the supersymmetric configurations and
we compute the equations of motion 
for the minimal area nonholomorphic 
curves. 
There exist some relevant works \cite{FU}-\cite{Ahnaug} 
along the line of \cite{ISS}.

\section{M-theory lift of symplectic gauge theory with massive flavors}

Let us describe the type IIA 
brane configuration of minimal energy supersymmetry breaking with
symplectic gauge group with massive flavors by taking magnetic theory
of supersymmetric brane configuration and moving the D6-branes to the
$(45)$ directions:there exist
$(N_f-N_c-2)$ ``color'' D4-branes stretched 
between D6-branes and an NS5-brane along the 
$x^6$ direction and tilted $(N_c+2)$ ``flavor'' D4-branes connecting to the
remaining D6-branes and NS5'-brane. 
In order to study symplectic or orthogonal gauge groups, we need to
add an additional orientifold 4-plane \cite{AOT,AOT1,CS,EGKRS,GK} 
with worldvolume $(01236)$
directions which is parallel to the above ``color'' D4-branes
and is not of finite extent in $x^6$ direction.
The spacetime reflection of this orientifold 4-plane(O4-plane) acts as
$(x^4,x^5,x^7,x^8,x^9) \rightarrow (-x^4,-x^5,-x^7,-x^8,-x^9)$ and therefore
each object which does not lie at the fixed points must have its
mirror image. One can see the change of the number of D4-branes(which determines the
rank of magnetic gauge group) during a moving of
NS5-brane across the D6-branes and NS5'-brane and there is a
contribution from the O4-plane charge. The intersection of two
different kinds of D4-branes above arises from the moving of D6-branes to
the $(45)$ directions.   
The various branes and O4-plane are located as follows: 

$\bullet$
One NS5-brane(colored by red)  with worldvolume $(012345)$ living at a point 
in the $(6789)$ directions. The positive piece in 45 directions has
its mirror image in the negative region in 45 directions.  

$\bullet$ 
One NS5'-brane(colored by blue) with worldvolume $(012389)$ living at a point 
in the $(4567)$ directions.  The positive piece in 89 directions has
its mirror image in the negative region in 89 directions. 

$\bullet$
$N_f$ D6-branes(dotted black) with worldvolume $(0123789)$ living at a point 
in the $(456)$ directions(i.e., positive in 45 directions and are located at
$x^6=0$) 
and its mirrors. 

$\bullet$
$(N_f-N_c-2)$ D4-branes(solid black) with worldvolume $(01236)$ living at a point 
in the $(45789)$ directions(positive in 45 directions) and its mirrors. 

$\bullet$ tilted
$(N_c+2)$ D4-branes(solid black) with worldvolume $(01236)$ living at a point 
in the $(45789)$ directions and its mirrors.

$\bullet$ 
One O4-plane(colored by green) with worldvolume $(01236)$ living at a point 
in the $(45789)$ directions. It is located at $x^4=x^5=x^8=x^9=0$. 

Now we draw this brane configuration which is nothing new and was
found in \cite{FGU} already and we repeat here 
for this paper to be self-complete as follows.

\begin{figure}[ht]
   \epsfxsize=5in 
\centerline{\epsffile{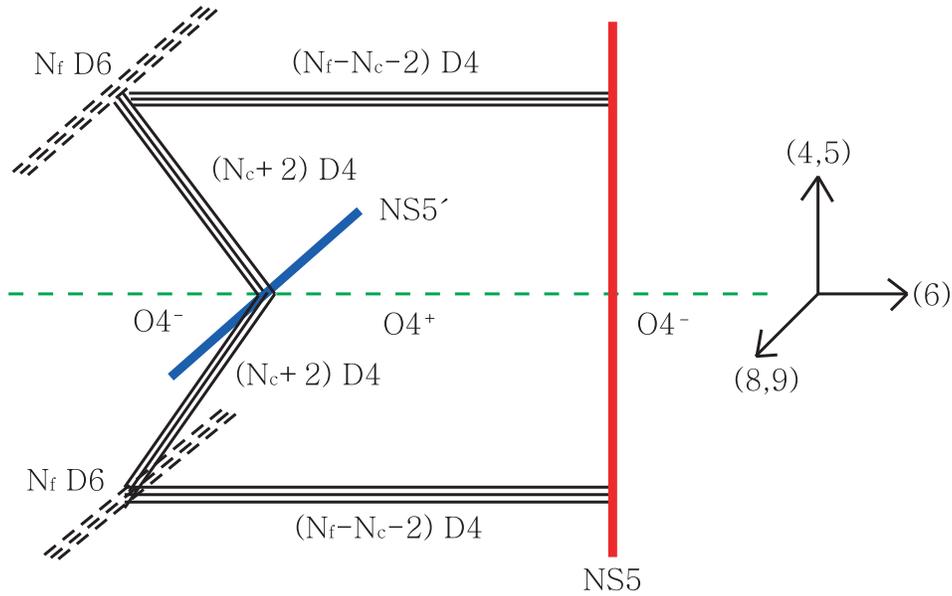}}
   \caption[FIG. \arabic{figure}.]{ 
The type IIA 
brane configuration of minimal energy supersymmetry breaking with
symplectic gauge group $Sp(N_f-N_c-2)$ with $2N_f$ massive flavors. 
For simplicity, we take
equal flavor masses(D6-branes have the same coordinates in 45
directions). 
The green dotted line stands for the
O4-plane. The O4-plane charge flips sign whenever one crosses a
D6-brane, NS5-brane, or NS5'-brane. The positive O4-plane charge
appears between NS5'-brane and NS5-brane \cite{GK}.  }
\label{fig1}
\end{figure}

Since the brane configuration gives only perturbative effect,
in order to get the non perturbative effect, we need to 
lift type IIA brane configuration to M-theory. 
The M5-brane configuration 
is in the 11-dimensional spacetime 
which is a direct product of flat seven-dimensional 
spacetime(0123789 directions) and the four-dimensional 
Taub-NUT space(456 and 10 directions).
The M5-branes 
span 0123 directions and
wrap on a Riemann surface inside transverse 45689 and 10 directions.
The metric by these six-dimensional transverse directions 
is given by \cite{Hawking}
\bea
ds_6^2 = G_{AB} dX^A dX^B = V d \vec{r}^2 + V^{-1}
\left( d x^{10} + \vec{\omega} \cdot d \vec{r} \right)^2
+ (dx^8)^2 + (dx^9)^2 
\label{metric}
\eea
where 456 directions are parametrized by $\vec{r}$, the harmonic function
$V$ sourced by the coincident $2N_f$ D6-branes located at
$x^4=x^5=x^6=0$ 
and the vector potential 
$\vec{\omega}$ are  given by 
\bea
V = 1+ \frac{2N_f R}{r}, \qquad \nabla \times \vec{\omega} =\nabla V.
\nonu
\eea
The compactification radius $R$ along 10th direction 
is given by a product of a string 
coupling $g_s$ and a string scale $\ell_s$ via $R=g_s \ell_s$.
Once the half of the D6-branes($N_f$ D6-branes) 
are located at $N_f$ different 45 directions with 
$x^6=0$, in general, 
then the remaining half of them(other $N_f$ D6-branes) 
can be located at the other side of 45 directions(reflected $N_f$ points)
with $x^6=0$ automatically via an orientifold O4-plane whose worldvolume
is the same as the one for D4-branes. See Figure 1.
 
On the other hand,
in the holomorphic coordinates \cite{Hori}, 
the transverse directions to the noncompact 0123 directions
are a product of a real line(7th direction) and a three-dimensional 
complex manifold(4568910 directions). 
Then  the Taub-NUT space is parametrized by 
two complex variables $(v,y)$ and flat two-dimensions are by
a complex variable $w$.
The mass dimensions of these variables are
given by 
\bea
[v]=1, \qquad [y]=4(N_c+1), \qquad [w]=2
\label{mass}
\eea
respectively. It is clear that 
the mass dimension of $v$ comes from the Seiberg-Witten 
curve for ${\cal N}=2$ 
$Sp(N_c)$ gauge theory with $2N_f$ matter fields.
For large $v$, since $w$ goes to $\mu v$ where 
$\mu$ is a mass of the adjoint chiral multiplet, the mass dimension of
$w$ is equal to 2.
The mass dimension of $y$ corresponding to a variable $\widetilde{t}$
in \cite{AOT}
can be determined by 
the boundary condition near 
$w =\infty$. See below the classification 2 characterized by 
NS' asymptotic region.

The explicit relations between 
the physical coordinates(4568910 directions) and 
holomorphic coordinates($v,y$ and $w$)
are given by
\bea
v =\frac{x^4 + i x^5}{\ell_s^2}, \qquad y =
\mu^{4(N_c+1)} e^{\frac{x^6-L_0+i x^{10}}{2R}} 
\left(\frac{r+x^6}{R}\right)^{N_f}, \qquad
w= \frac{x^8 + i x^9}{R \ell_s^2}.
\nonu
\eea
Note that in the brane configuration of \cite{BGHSS}, the location of
NS5'-brane is given by $v=0$ and $x^6=L_0$.
We insert this normalization constant $L_0$ above.
The power of dimensionful scale 
$\mu$ in $y$ indicates the correct mass dimension above (\ref{mass}).
The 11-dimensional Planck scale $\ell_p$ has a relation 
$\ell_p^3 = R \ell_s^2$ and the product of constant terms in $v$ and $w$
gives $\frac{R}{\ell_p^6}$ which is related to the M5-brane tension.
The $N_f$-dependent term in harmonic function $V$ enduces 
the $N_f$-dependent term in $y$ and the 
precise expression $\left(\frac{r+x^6}{R}\right)$ will appear also
after we solve the function
$g(s)$ later using the equations of motion.

The supersymmetric M5-brane 
configurations for massless matter are described by 
two holomorphic curves \cite{AOT} in the IR free magnetic range of 
$N_c+2 < N_f < \frac{3}{2} (N_c+1)$ \cite{ISS}
as follows:
\bea
{\cal C}_{NS} &:&
w(z) = 0, \qquad v(z)=z, \qquad y(z) =  \Lambda_{N=1}^{6(N_c+1)-2N_f} 
z^{2(N_f-N_c-1)}
\nonu \\
{\cal C}_{NS'} & : &
w(z) = z, \qquad v(z) =0, \qquad y(z) = z^{2(N_c+1)}
\nonu
\eea
where the ${\cal C}_{NS}$ component of the curve
describes the $(N_f-N_c-2)$ ``color'' D4-branes ending on the NS5-brane 
and ${\cal C}_{NS'}$ component of the curve describes
the $(N_c+2)$ ``flavor'' D4-branes ending on the NS5'-brane \cite{FGU}.
Of course, their mirrors, $(N_f-N_c-2)$ D4-branes and $(N_c+2)$ D4-branes
exist as a reflection via an orientifold O4-plane.
See Figure 1.
We follow the convention for the orientifold O4-plane as \cite{EGKRS} which 
is different from what used in \cite{FGU}.
One can easily see that 
the mass dimensions of (\ref{mass}) hold for these curves. 
Since $v$ has a charge $(2,0)$ under, 
the rotations in the 45 directions and 89 directions,
the $U(1)_{45} \times U(1)_{89}$
and $w$ has a charge $(0,2)$ under the $U(1)_{45} \times U(1)_{89}$,
one can read off the charges for $y$ and $\Lambda_{N=1}^{6(N_c+1)-2N_f}$.
The former has a charge $(0, 4N_c+4)$ and the latter has a charge
$(4N_c+4-4N_f, 4N_c+4)$ under the $U(1)_{45} \times U(1)_{89}$.

When the quarks have equal mass $m_f$(i.e., D6-branes are located at one fixed
value for $v=v_{m_f}$ and its mirrors at $v=-v_{m_f}$), 
the supersymmetric curve for $\mu \rightarrow \infty$ 
has only one component \cite{AOT} 
and is given by
\bea
w(z)  = z, \qquad v(z) = 2^{\frac{N_f-2N_c}{N_c+1}} m_f \frac{z_0}{z}, 
\qquad y(z) = 
z^{2(N_c+1-N_f)} \left( z^2 -z_0^2 \right)^{N_f},
\label{onecomp}
\eea
where
$
z_0^{N_c+1} \equiv 2^{2N_c-N_f} m_f^{N_f-N_c-1} 
\Lambda^{3(N_c+1)-N_f}_{N=1} 
$ and $v$ goes to zero for large $w$ and $v$ approaches 
$\pm 2^{\frac{N_f-2N_c}{N_c+1}} m_f$ as $w$ goes to $\pm z_0$.
One can compute the energy of the supersymmetry breaking vacuum,
$V_0$, using the effective field theory or the DBI action(i.e., the length
of D4-branes). The former 
can be written as \cite{BGHSS,ISS}
\bea
V_0 = \frac{2(N_c+2) |\Delta x|^2}{g_s \ell_s^5 L_0}
\label{V}
\eea
and the latter can be written, by noting that there exist
$(N_f-N_c-2)$ D4-branes between D6-branes and NS5-brane in Figure 7 of
\cite{BGHSS} or in Figure 1 and its mirrors and $(N_c+2)$ D4-branes between D6-branes
and NS5'-brane and its mirrors, as
\bea
V_{DBI} =\frac{1}{g_s \ell_s^5} 
\left(2(N_c+2) \sqrt{|\Delta x|^2+L_0^2} + 
2(N_f-N_c-2)(L_0 + |\Delta L|) \right)
\label{V1}
\eea
by summing up all the contributions when we take the product of
D4-brane tension and the length of D4-brane.
In the limit of $|\Delta x| << L_0$, the $|\Delta x|$-dependent part 
of (\ref{V}) and (\ref{V1}) agrees with each other.
The energy difference between the tachyonic state and the vacuum  is given by
\bea
\Delta V_{DBI} = V_{tach}-V_{DBI}=
\frac{1}{g_s \ell_s^5} 2(N_f-N_c-2)
\left( \sqrt{|\Delta x|^2+L_0^2} -L_0 \right)
\nonu
\eea
which is the same as the energy difference coming from field theory 
in the limit  $|\Delta x| << L_0$ and  
$V_{tach}=\frac{1}{g_s \ell_s^5} 
(2N_f \sqrt{|\Delta x|^2+L_0^2} + 
2(N_f-N_c-2)|\Delta L|)$ which can be obtained from Figure 6 of
\cite{BGHSS} as follows: $(N_f-N_c-2)$ D4-branes between NS5'-brane and
NS5-brane and $N_f$ D4-branes between D6-branes and NS5'-brane as well
as their mirrors.

Now one can 
see the behavior of the supersymmetric curves \cite{AOT}
in various limits as follows:

1. $v \rightarrow \infty$ limit implies
\bea
w \rightarrow 0, \qquad y \rightarrow    \Lambda_{N=1}^{6(N_c+1)-2N_f} 
v^{2(N_f-N_c-1)} + \cdots \qquad
\mbox{NS asymptotic region}   
\nonu
\eea

2.  $w \rightarrow \infty$ limit implies
\bea
v \rightarrow   m_f, \qquad y \rightarrow w^{2(N_c+1)} +\cdots
\qquad \mbox{NS' asymptotic region}
\nonu
\eea
Note that the $v$ here is different from $v$ of \cite{AOT} by 
a constant shift $m_f$. 
Instead of moving D6-branes in the $v$ direction,
the NS5'-brane is moving in the $v$ direction holding 
everything else fixed.
The mirrors of ``color'' and ``flavor'' D4-branes are displaced 
appropriately.
Note that in Figure 1, the origin for 45 directions is located at the
position of NS5-brane while in the coordinate we use in the above NS'
asymptotic region, it is located at the position of mirror $N_f$ D6-branes.
Therefore, the location of original $N_f$ D6-branes is given by $2m_f$.  

3. The map between the holomorphic and physical coordinates
requires 
the condition \cite{BGHSS}
\bea
 y=0 \qquad \mbox{only if} \qquad v=0.
\nonu
\eea

For the holomorphic massless curve one can see the two brane
configurations, one for electric case and the other for 
magnetic case \cite{AOT,CS}.
In other words, the emergence of the dual $Sp(N_f-N_c-2)$
gauge group arises by looking at the same M-theory M5-brane in two
different limits.
However, for nonzero mass, the only one component M-theory curve 
(\ref{onecomp})
reduces to the 
supersymmetric electric brane configuration 
as $R \rightarrow 0$.

The dependence of $x^{10}$ in the curve \cite{BGHSS}
comes from the phase of $w$ as follows:
\bea
\mbox{arg} \left( x^8 + i x^9 \right) =\frac{x^{10}}{4(N_c+1)R}.
\nonu
\eea
Here the $U(1)_{89}$ rotation by $\frac{\pi}{2(N_c+1)}$ leaves 
the M-theory curve invariant.
In particular, the charges of $\Lambda_{N=1}^{6N_c+6-2N_f}$ imply
that the groups $U(1)_{45}$ and $U(1)_{89}$ are broken to 
their discrete subgroups ${\bf Z}_{4(N_c+1-N_f)}$ and ${\bf Z}_{4(N_c+1)}$ 
respectively \cite{AOT,CS}.
In the brane configuration, the above $U(1)_R$ is identified with
rotations in the 89 plane and the anomaly of the corresponding $U(1)_R$
symmetry group is identified as a shift in $x^{10}$ under
rotations in the 89 plane \cite{Witten}. 

Now it is ready to consider M5-brane wrapping 
a non holomorphic mimimal area surface 
in the Taub-NUT mutliplied by flat two dimensions 
satisfying the boundary conditions above \cite{BGHSS}.
The solution for non-holomorphic curve  can be written as
\bea
x^4 = f(s), \qquad x^5 =0, \qquad 
x^8 + i x^9 = e^{i \frac{x^{10}}{4(N_c+1) R}} g(s), \qquad
x^6 =s.
\label{solution}
\eea
We would like to solve the equations of motion for unknown functions
$f(s)$ and $g(s)$.
The action of a surface parametrized by $x^6$ and $x^{10}$ is 
given by
\bea
A = \int d^2 z \sqrt{\mbox{det} G_{AB}
\pa_a X^A \pa_b X^B }
\label{action}
\eea
where the six-dimensional metric $G_{AB}$ is given by
(\ref{metric}).
Then the action, by inserting the solution (\ref{solution}) 
and (\ref{metric}) into 
the action (\ref{action}), can be written as
\bea
A =\int ds \sqrt{\left[ V^{-1} +\frac{g^2}{16 R^2(N_c+1)^2 } 
\right] \left[ V(1+f^{'2}) + g^{'2}\right]}
\nonu
\eea
where the harmonic function becomes by realizing that the original
$N_f$ D6-branes are located at $v=2 \Delta x/\ell_s^2$(and $x^6=0$) and its mirrors are
located at $v=0$(and $x^6=0$)
\bea
V = 1 +\frac{N_f R}{\sqrt{(f-2\Delta x)^2 + s^2}} + \frac{N_f R}{\sqrt{f^2 + s^2}}.
\label{harmonic}
\eea
Compared with the case for $SU(N_c)$ gauge theory \cite{BGHSS},
the Lagrangian doesn't have any big difference in the sense that 
the 
color- and flavor-dependent terms are a little bit different. 
Then, one can follow similar procedure of \cite{BGHSS} and arrives
at an exact expression for the function $f(s)$ with modified 
$N_c$- and $N_f$-dependent terms. However, 
when  we look at (A.4) of \cite{BGHSS}, it turns out
that since it does not 
depend on $N_c$ or $N_f$, the solution for $f(s)$ for symplectic group
case is the same as the one in $SU(N_c)$ case. 
There exsits always a straight line solution  
$f''(s)=0$ as in \cite{BGHSS}.

When 
\bea
f(s) =  \Delta x
\nonu
\eea
which satisfies $f''(s)=0$, 
then the equation (A.3) of \cite{BGHSS}
implies that $g'(s) =\frac{V}{4(N_c+1) R} g(s)$ with $V(s)=
1 + \frac{2N_f R}{\sqrt{(\Delta x)^2 + s^2}}
$.
So this first order differential equation provides the following solution
\bea
g(s) = R \ell_s^2 \mu^2 e^{\frac{s-L_0}{4(N_c+1)R}} 
\left( \frac{s+ \sqrt{(\Delta x)^2 + s^2}}{R} \right)^{
\frac{N_f}{2(N_c+1)}}.
\label{sol}
\eea
The integration constant which is $s$-independent term 
is fixed by the boundary condition
where $y \rightarrow w^{2N_c +2}$ we have discussed before and is given by
$ R \ell_s^2 \mu^2 e^{-\frac{L_0}{4(N_c+1)R}} 
\left( \frac{1}{R} \right)^{
\frac{N_f}{2(N_c+1)}}$. In other words, 
this is a simple solution $v=\frac{ \Delta x}{\ell_s^2} 
= m_f$ and  $y=w^{2(N_c+1)}$.
Even if $\Delta x$ is equal to zero, the function $g(s)$ doesn't vanish 
implying
that both $w$ and $y$ don't vanish. 
Therefore, this doesn't lead to the classification 3
above: it does not end on D6-brane.
Therefore, there is no smooth non-holomorphic M-theory curve.
  
Instead of imposing the boundary condition 
at large $s$, we require that M5-branes end on the D6-branes: 
$f(s)$ at $s=0$ vanishes.
For the case of
\bea
f(s) = c s
\nonu
\eea
which satisfies $f''(s)=0$,
the (A.3) of \cite{BGHSS} implies that there exists 
$
g'(s) = \frac{\sqrt{1+f'(s)^2}}{4(N_c+1) R} V g(s)
$
and when $L=L_0$ and $c=\frac{ \Delta x}{L_0}$, 
this straight line solution 
will lead to the type IIA brane configuration \cite{FGU} 
in the $R \rightarrow 0$ 
limit. 
However, the behavior at infinity is different from the above 
classification 1 and 2.
Therefore, the supersymmetric brane configuration and the supersymmetry 
breaking brane configuration are vacua of different theories because
the boundary conditions at infinity are different.

Also the analysis for a kink quasi-solution \cite{BGHSS},  in order to understand 
the absence of meta-stable vacuum in the ``D-brane limit'' of MQCD,
can be done similarly. 
For given expression of (\ref{sol}) corresponding to $x^6 > L$, 
the magnitude of $w$ and $v$ are given by
\bea
x^6 < L:
\quad |x^8 + i x^9| & = & R \ell_s^2 \mu^2 e^{-\frac{(L_0-L)}{4(N_c+1)R}}
e^{\frac{\sqrt{1+(\frac{\Delta x}{L})^2}
\left(x^6-L\right)}{4(N_c+1)R}}
\left( \frac{
x^6+\sqrt{(x^6)^2+(\frac{\Delta x }{L} x^6)^2}}{R} 
\right)^{\frac{N_f}{2(N_c+1)}}
\nonu \\
 x^4 & = &  \frac{\Delta x}{L} x^6,
\nonu \\
x^6 > L:
\quad |x^8 + i x^9| & = & R \ell_s^2 \mu^2 e^{\frac{(x^6-L_0)}{4(N_c+1)R}}
\left( \frac{
x^6+\sqrt{(x^6)^2+(\Delta x)^2}}{R} \right)^{\frac{N_f}{2(N_c+1)}}
\nonu \\
 x^4 & = &  \Delta x. 
\nonu
\eea
Plugging these into the action,
one gets the potential that has runaway behavior as for $SU(N_c)$
case \cite{BGHSS} which does not depend on $N_c$ or $N_f$ and
that is proportional to the energy difference 
$\Delta V_{DBI}$ with a replacement $L_0$ by $L$ 
we have considered before.


Let us make some comments on the solution for the differential equation 
satisfying (A.4) in \cite{BGHSS}. Also in our case, we have the same
differential equation as (A.4) of \cite{BGHSS} because we didn't put
the form of the potential (\ref{harmonic}) in arriving at (A.4).  
In \cite{BGHSS}, they have tried to search for the possibility for the
other solutions with the right boundary conditions by substituting the
explicit form for the potential and have obtained nonlinear
differential equation (A.5) of
\cite{BGHSS}. Miraculously, the exact solution for the $f(s)$ was
found through (A.5)-(A.10) of \cite{BGHSS}. However, since our potential has
an extra piece in (\ref{harmonic}), the third order nonlinear
differential equation for $f(s)$ cannot be solved exactly. The above
solutions $f(s) =\Delta x$ and $f(s) = c s$ are particular solutions
and, in principle, there could exist a solution 
having the correct boundary conditions both at infinity 
and at the D6-brane with $f''(s) \neq 0$. It seems to be difficult to
construct this solution even if one uses the numerical analysis for
the complicated diffferential equation.

In summary, 
we have considered the lifting of supersymmetry breaking brane 
configurations of \cite{ISS,FGU} for symplectic gauge group with 
massive flavors. The main object was M5-branes wrapping nonholomorphic
minimal area curves in Taub-NUT times flat two dimensions. So 
we have found equations of motion for the ansatz characterized by
(\ref{solution}) with the same boundary conditions as supersymmetric 
vacua \cite{AOT,CS}
and it turned out there was no meta-stable brane configuration in the
``D-brane limit'' of MQCD (at least for the above particular solutions 
$f(s) =\Delta x$ and $f(s) = cs$), like as $SU(N_c)$ gauge group 
case \cite{BGHSS}.   

\section{M-theory lift of orthogonal gauge theory with massive flavors}

As for previous section, the type IIA 
brane configuration of minimal energy supersymmetry breaking with
orthogonal gauge group with massive flavors can be understood similarly:
$(N_f-\frac{N_c}{2}+2)$ ``color'' D4-branes stretched 
between D6-branes and an NS5-brane along the 
$x^6$ direction and tilted $(\frac{N_c}{2}-2)$ ``flavor'' D4-branes connecting to the
remaining D6-branes and NS5'-brane. 
We focus on even $N_c$ case.
For odd $N_c$, an extra single D4-brane is stuck at
the O4-plane and this extra D4-brane 
is not affected by a moving of D6-branes to 45 directions.
One can see the change of the number of D4-branes(which determines the
rank of magnetic gauge group) during a moving of
NS5-brane across the D6-branes and NS5'-brane and there is a
contribution from the O4-plane charge which is different from the one
for symplectic gauge group. The intersection of two
different kinds of D4-branes above arises from the moving of D6-branes to
the 45 directions.   
The various branes and O4-plane are located as follows: 

$\bullet$
One NS5-brane(colored by red)  with worldvolume $(012345)$ living at a point 
in the $(6789)$ directions.  

$\bullet$ 
One NS5'-brane(colored by blue) with worldvolume $(012389)$ living at a point 
in the $(4567)$ directions.  

$\bullet$
$N_f$ D6-branes(dotted black) with worldvolume $(0123789)$ living at a point 
in the $(456)$ directions(i.e., positive in 45 directions and are located at
$x^6=0$) 
and its mirrors. 

$\bullet$
$(N_f-\frac{N_c}{2}+2)$ D4-branes(solid black) with worldvolume $(01236)$ living at a point 
in the $(45789)$ directions(positive in 45 directions) and its mirrors. 

$\bullet$ tilted
$(\frac{N_c}{2}-2)$ D4-branes(solid black) with worldvolume $(01236)$ living at a point 
in the $(45789)$ directions and its mirrors.

$\bullet$ 
One O4-plane(colored by green) with worldvolume $(01236)$ living at a point 
in the $(45789)$ directions. It is located at $x^4=x^5=x^8=x^9=0$. 

Now we repeat the brane configuration here 
for this paper to be self-complete as follows.

\begin{figure}[ht]
   \epsfxsize=5in 
\centerline{\epsffile{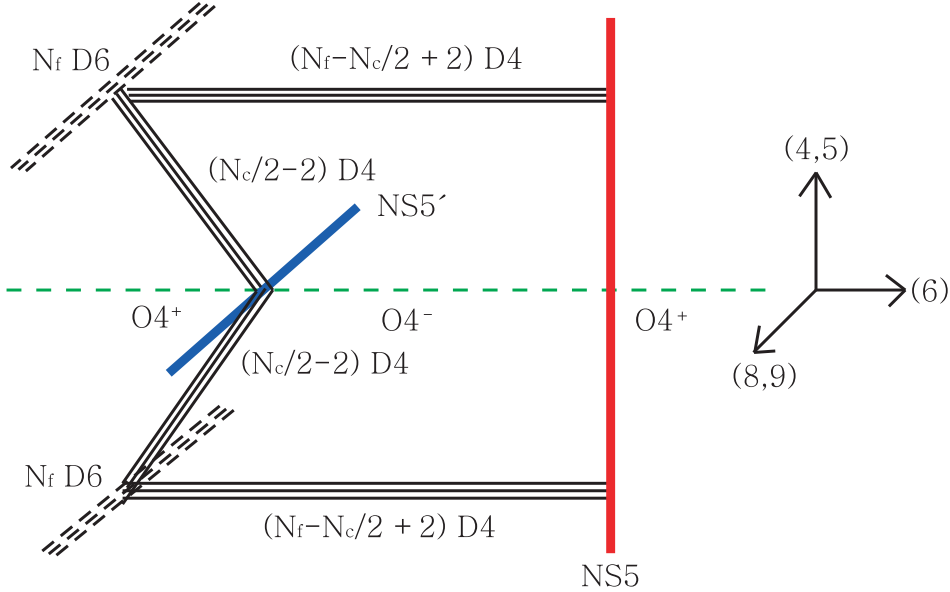}}
   \caption[FIG. \arabic{figure}.]{ 
The type IIA 
brane configuration of minimal energy supersymmetry breaking with
orthogonal gauge group $SO(2N_f-N_c+4)$ with $2N_f$ massive flavors. For simplicity, we take
equal flavor masses and even number of $N_c$. 
For odd $N_c$, an extra single D4-brane is stuck at
the O4-plane denoted by green dotted line. 
 The negative O4-plane charge
appears between NS5'-brane and NS5-brane.
The O4-plane charge flips sign whenever one crosses a
D6-brane, NS5-brane, or NS5'-brane. }
\label{fig2}
\end{figure}

In this section, we can proceed for orthogonal gauge group based on
the previous analysis and let us  
describe the main facts.
The mass dimensions of complex variables are
given by 
\bea
[v]=1, \qquad [y]=2(N_c-2), \qquad [w]=2.
\label{mass1}
\eea
The mass dimension of $v$ comes from the Seiberg-Witten 
curve for ${\cal N}=2$ 
$SO(N_c)$ gauge theory with $2N_f$ matter fields.
For large $v$, since the $w$ behaves as $\mu v$, 
its mass dimension is equal to 2.
The mass dimension of $y$ corresponding to a variable $\widetilde{t}$
in \cite{AOT1}
can be determined by 
the boundary condition near 
$w =\infty$. See below the classification 2 characterized by 
NS' asymptotic region.
The holomorphic coordinates($v,y$ and $w$)
are given by
\bea
v =\frac{x^4 + i x^5}{\ell_s^2}, \qquad y =
\mu^{2(N_c-2)} e^{\frac{x^6-L_0+i x^{10}}{2R}} 
\left(\frac{r+x^6}{R}\right)^{N_f}, \qquad
w= \frac{x^8 + i x^9}{R \ell_s^2}.
\nonu
\eea
The power of dimensionful scale 
$\mu$ in $y$ indicates the correct mass dimension above (\ref{mass1}).

The supersymmetric M5-brane 
configurations for massless matter are described by 
two holomorphic curves \cite{AOT1} in the IR free magnetic range of 
$N_c-4 < 2N_f < \frac{3}{2} (N_c-2)$ \cite{ISS}
as follows:
\bea
{\cal C}_{NS} &:&
w(z) = 0, \qquad v(z)=z, \qquad y(z) =  \Lambda_{N=1}^{3N_c-6-2N_f} 
z^{2N_f-N_c+2}
\nonu \\
{\cal C}_{NS'} & : &
w(z) = z, \qquad v(z) =0, \qquad y(z) = z^{N_c-2}
\nonu
\eea
where the ${\cal C}_{NS}$ component of the curve
describes the $(2N_f-N_c+4)$ D4-branes ending on the NS5-brane 
and ${\cal C}_{NS'}$ component of the curve describes
the $(N_c-4)$ D4-branes ending on the NS5'-brane.
Note that these are total number of D4-branes including their mirrors.
For even $N_c$, one can imagine half of the  total 
``color'' $(2N_f-N_c+4)$ D4-branes are located at a fixed value of $v$ 
above an orientifold O4-plane and its mirrors are located below 
an orientifold O4-plane. For odd $N_c$, a single D4-brane is stuck at
the orientifold O4-plane and half of the remaining D4-branes and its mirrors
are located like as for the even $N_c$ case \cite{EGKRS}. Also this 
consideration of the locations of D4-branes holds for 
``flavor'' $(N_c-4)$ D4-branes.
See Figure 2.
The charges of $y$ and $\Lambda_{N=1}^{3N_c-6-2N_f}$ are $(0,2N_c-4)$ and 
$(2N_c-4-4N_f,2N_c-4)$
under the $U(1)_{45} \times U(1)_{89}$ respectively which can be
determined as previous section without any difficulty.

The behavior of the supersymmetric curves \cite{AOT1}
in various limits can be summarized as follows:

1. $v \rightarrow \infty$ limit implies
\bea
w \rightarrow 0, \qquad y \rightarrow    \Lambda_{N=1}^{3N_c-6-2N_f} 
v^{2N_f-N_c+2} + \cdots \qquad
\mbox{NS asymptotic region}   
\nonu
\eea

2.  $w \rightarrow \infty$ limit implies
\bea
v \rightarrow  m_f, \qquad y \rightarrow w^{N_c-2} +\cdots
\qquad \mbox{NS' asymptotic region}
\nonu
\eea

The energy of the supersymmetry breaking vacuum $V_0$
from the field theory side 
can be written as \cite{ISS} with appropriate normalization \cite{ISS}
\bea
V_0 = \frac{(N_c-4) |\Delta x|^2}{g_s \ell_s^5 L_0}
\nonu 
\eea
and that from the DBI action can be written as
\bea
V_{DBI} =\frac{1}{g_s \ell_s^5} 
\left((N_c-4) \sqrt{|\Delta x|^2+L_0^2} + 
(2N_f-N_c+4)(L_0 + |\Delta L|) \right)
\nonu
\eea
by realizing that there are $(2N_f-N_c+4)$ color D4-branes and 
$(N_c-4)$ flavor D4-branes altogether in Figure 7 of \cite{BGHSS} or
Figure 2 and
summing up all the contributions.
In the limit of $|\Delta x| << L_0$, the $|\Delta x|$-dependent part
for the energies  
agrees with each other.
The energy difference between the tachyonic state and the vacuum is given by
\bea
\Delta V_{DBI} = V_{tach}-V_{DBI}= \frac{1}{g_s \ell_s^5} (2N_f-N_c+4)
\left( \sqrt{|\Delta x|^2+L_0^2} -L_0 \right)
\nonu
\eea
which is the same as the energy difference coming from field theory 
in the limit  $|\Delta x| << L_0$
and $V_{tach}=\frac{1}{g_s \ell_s^5} 
(2N_f \sqrt{|\Delta x|^2+L_0^2} + 
(2N_f-N_c+4)|\Delta L|)$ where there exist $2N_f$ flavor
D4-branes and $(2N_f-N_c+4)$ color D4-branes in Figure 6 of \cite{BGHSS}.

3. The map between the holomorphic and physical coordinates
requires 
the condition \cite{BGHSS}
\bea
 y=0 \qquad \mbox{only if} \qquad v=0.
\nonu
\eea

For the holomorphic massless curve one can see the two brane
configurations, one for electric case and the other for 
magnetic case \cite{AOT1,CS}.
In other words, one sees the emergence of the dual $SO(2N_f-N_c+4)$
gauge group by looking at the same M-theory M5-brane in two
different limits.
However, for nonzero mass, the only one component M-theory curve 
reduces to the 
supersymmetric electric brane configuration 
as $R \rightarrow 0$.

The dependence of $x^{10}$ in the curve \cite{BGHSS}
comes from the phase of $w$ as follows:
\bea
\mbox{arg} \left( x^8 + i x^9 \right) =\frac{x^{10}}{2(N_c-2)R}.
\nonu
\eea
In particular, as in previous case, 
the charges of $\Lambda_{N=1}^{3N_c-6-2N_f}$ imply
that the groups $U(1)_{45}$ and $U(1)_{89}$ are broken to 
their discrete subgroups ${\bf Z}_{2(N_c-2-2N_f)}$ and ${\bf Z}_{2(N_c-2)}$ 
respectively \cite{AOT1}.

The solution for non-holomorphic curve  can be written as
\bea
x^4 = f(s), \qquad x^5 =0, \qquad 
x^8 + i x^9 = e^{i \frac{x^{10}}{2(N_c-2) R}} g(s), \qquad
x^6 =s.
\label{solution1}
\eea
By inserting the solution (\ref{solution1}) 
and (\ref{metric}) into 
the action (\ref{action}), the action can be written as
\bea
A =\int ds \sqrt{\left[ V^{-1} +\frac{g^2}{4 R^2(N_c-2)^2 } 
\right] \left[ V(1+f^{'2}) + g^{'2}\right]}
\nonu
\eea
which is almost the same form as previous section and
only the color-dependent term is different. 
The harmonic function is the
same as before and is given by
(\ref{harmonic}).

When $f(s) =  \Delta x$, 
the first order differential equation provides the following solution
\bea
g(s) = R \ell_s^2 \mu^2 e^{\frac{s-L_0}{2(N_c-2)R}} 
\left( \frac{s+ \sqrt{(\Delta x)^2 + s^2}}{R} \right)^{
\frac{N_f}{(N_c-2)}}.
\nonu
\eea
This is a simple solution $v=\frac{ \Delta x}{\ell_s^2} 
= m_f$ and  $y=w^{N_c-2}$.
Even if $\Delta x$ is equal to zero, the function $g(s)$ doesn't vanish 
implying
that both $w$ and $y$ don't vanish. 
There is no smooth non-holomorphic M-theory curve.
On the other hand, for the case of
$
f(s) = c s$,
it is easy to see that 
when $L=L_0$ and $c=\frac{ \Delta x}{L_0}$, 
this straight line solution 
will lead to the type IIA brane configuration \cite{FGU} 
in the $R \rightarrow 0$ 
limit. However, the behavior at infinity is different from the above 
classification 1 and 2.

Also the analysis for a kink quasi-solution,  in order to understand 
the absence of meta-stable vacuum in the ``D-brane limit'' of MQCD,
can be done similarly as in previous case. 

In summary, 
we have described the lifting of supersymmetry breaking brane 
configurations of \cite{ISS,FGU} for orthogonal gauge group with 
massive flavors.  
We have found equations of motion for the ansatz characterized by
(\ref{solution1}) with the same boundary conditions as supersymmetric 
vacua \cite{AOT1,CS}. 
There was no meta-stable brane configuration in the
``D-brane limit'' of MQCD, like as $SU(N_c)$ gauge group 
case \cite{BGHSS} or $Sp(N_c)$ gauge group case in previous section.   


\vspace{.7cm}

\centerline{\bf Acknowledgments}

We would like to thank David Shih  for discussions and providing the
transparencies of his talks on \cite{BGHSS}.
This work was supported by grant No.
R01-2006-000-10965-0 from the Basic Research Program of the Korea
Science \& Engineering Foundation. 

\end{document}